\documentstyle[aps,prd,eqsecnum,epsf]{revtex}

\def\bm{\bbox}

\begin{document}
\thispagestyle{empty}

%
%
\leftline{\epsfbox{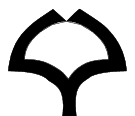}}
\vspace{-10.0mm}
{\baselineskip-4pt
\font\yitp=cmmib10 scaled\magstep2
\font\elevenmib=cmmib10 scaled\magstep1  \skewchar\elevenmib='177
\leftline{\baselineskip20pt
\hspace{12mm} 
\vbox to0pt
   { {\yitp\hbox{Osaka \hspace{1.5mm} University} }
     {\large\sl\hbox{{Theoretical Astrophysics}} }\vss}}

%
%
{\baselineskip0pt
\rightline{\large\baselineskip14pt\rm\vbox
        to20pt{\hbox{April 2001}
               \hbox{OU-TAP-162}
\vss}}
}
\vskip15mm

\begin{center}
{\Large\bf Quantum fluctuations in\\
brane-world inflation without
an inflaton on the brane}
\end{center}

\begin{center}
{\large Norichika Sago, Yoshiaki Himemoto and Misao Sasaki}\\
\vskip 1mm
\sl{Department of Earth and Space Science,
Graduate School of Science,\\
Osaka University, Toyonaka 560-0043, Japan}
\end{center}

\begin{abstract}
A Randall-Sundrum type brane-cosmological model in which slow-roll
inflation on the brane is driven solely by a bulk scalar field was
recently proposed by Himemoto and Sasaki.
We analyze their model in detail and
calculate the quantum fluctuations of the bulk scalar field
$\phi$ with $m^2=V''(\phi)$. We decompose the bulk scalar field
into the infinite mass spectrum of 4-dimensional fields;
the field with the smallest mass square, called the zero mode,
and the Kaluza-Klein modes above it with a mass gap.
We find the zero-mode dominance of the classical solution
holds if $|m^2|\bar\ell^2\ll1$,
where $\bar{\ell}$ is the curvature radius of the effectively
anti-de Sitter bulk, but it is violated if $|m^2|\bar\ell^2\gg1$,
 though the violation is very small.
Then we evaluate the vacuum expectation value
$\langle\delta\phi^2\rangle$ on the brane.
We find the zero-mode contribution completely dominates
if $|m^2|\bar{\ell}^2\ll 1$ similar to the case of classical background.
In contrast, we find the Kaluza-Klein contribution is small
but non-negligible if the value of $|m^2|\bar{\ell}^2$ is large.
\end{abstract}



\section{Introduction}
Recent progress in string theory suggests that our space-time is not
4-dimensional but higher dimensional.
Horava and Witten showed that a desirable gauge theory appears on the
10-dimensional boundaries of the 11-dimensional space-time
$M^{10}\times S^1/Z_2$ \cite{Horava:1996qa,Horava:1996ma}.
This opened up the possibility that we may live on a brane embedded
in a higher dimensional spacetime.
Arkani-Hamed {\it et al}. \cite{Arkani-Hamed}
investigated such a possibility and showed that the size of
the extra dimensions may be as large as 1 mm.
But gravity was not fully taken into account in their work.
Then taking the gravity into account, Randall and Sundrum showed
the existence of 5-dimensional models that realize the Horava-Witten theory
if the 5-dimensional cosmological constant is
negative \cite{Randall:1999ee,Randall:1999vf}.

In their first paper \cite{Randall:1999ee},
Randall and Sundrum
found that two flat 4-dimensional Minkowski branes can be naturally
embedded in the 5-dimensional anti-de Sitter (AdS$_5$) space
for appropriately tuned values of the
brane tension; $\sigma=\pm\sqrt{-6\Lambda_5}/\kappa_5^2$
where $\Lambda_5$ ($<0$) is the 5-dimensional cosmological constant
and $\kappa_5^2$ is the 5-dimensional gravitational constant.
They argued that the mass-hierarchy problem may be solved if we
live on the negative tension brane. However, it was shown that there
exists the so-called radion mode that describes the relative
motion between the two branes and this mode causes an unacceptable
modification of the effective gravity on the negative tension
brane \cite{Garriga:2000yh,Charmousis:2000rg} unless the radion is
somehow stabilized \cite{Tanaka:2000er}.

In their second paper \cite{Randall:1999vf}, they investigated the case of
a single positive tension brane in the AdS$_5$ bulk
and showed that gravity is confined within the curvature radius
$\ell=\sqrt{6/(-\Lambda_5)}$ of the AdS$_5$ and
the 4-dimensional Einstein gravity is recovered
on the brane on scales greater than $\ell$ despite the fact that
the extra dimension is infinite.
This raised an intriguing possibility that our universe is indeed
a brane world in a higher-dimensional spacetime, and
 a lot of attention has been paid to the cosmology
of a Randall-Sundrum type brane world
\cite{Shiromizu:2000wj,Mukohyama:2000qx,Kraus:1999it,Garriga:2000bq,%
Ida:2000ui,Kim:2000ta,vandeBruck:2000ha,Maartens:2000fg,Kodama:2000fa,%
Langlois:2000ia,vandeBruck:2000ju,Koyama:2000cc,Anchordoqui,%
Chamblin:1999ya,Kaloper:1999sm,Nihei:1999mt,%
Kim:2000ja,Hawking,Nojiri,Maartens:2000hf,Langlois:2001iu,%
Gen:2000nu,Langlois:2000ns,Yokoyama:2001nw,%
Kobayashi:2001yh,Gilkey:2001mj,Himemoto:2001nd}.
Among these studies of brane-world cosmology,
Himemoto and Sasaki recently investigated
the possibility of brane-world inflation induced by a
dynamical bulk scalar field without introducing
an inflaton field on the brane \cite{Himemoto:2001nd}.
The reheating after inflation based on this scenario has
been discussed by Yokoyama and Himemoto \cite{Yokoyama:2001nw}.
Possible realizations of such brane-world inflation models
have been discussed in \cite{Anchordoqui,Hawking,Nojiri}.

In \cite{Himemoto:2001nd}, Himemoto and Sasaki found a perturbative solution
of the bulk scalar field in the effective AdS$_5$ background (with a
modified curvature radius $\bar\ell$ larger than the original $\ell$)
and showed that slow-roll inflation
is realized on the brane if $|m^2|\ll H^2$, where $m$
is the mass of the bulk scalar field and $H$ is the Hubble
parameter on the brane, irrespective of the value of $m^2\bar\ell^2$.
This result is interesting in that the dynamics of
inflation on the brane is unaffected by the curvature scale of the bulk.
Naively one would expect otherwise if $|m^2|\bar\ell^2\gg1$,
since this implies $H^2\bar\ell^2\gg 1$ for $|m^2|\ll H^2$,
which is the case when the gravity on the brane on the Hubble horizon
scale is significantly affected.

In this paper, taking up the brane-inflation scenario proposed
in \cite{Himemoto:2001nd},
we discuss the $m^2$ dependence of the classical background more carefully
and investigate the effect of quantum fluctuations of
the bulk scalar field on the dynamics of the brane.
In particular, we are interested in whether there
appears a significant effect on the brane if $|m^2|\bar\ell^2\gg1$.
Recently Kobayashi, Koyama and Soda \cite{Kobayashi:2001yh}
have considered the quantum fluctuations of a
massless bulk scalar field $\phi$ in the AdS$_5$ background.
The bulk field $\phi$ may be decomposed into the zero mode
and the Kaluza-Klein modes. The former
corresponds to a massless 4-dimensional field and the
latter to massive 4-dimensional fields.
They evaluated the contribution of
the Kaluza-Klein modes to $\langle\phi^2\rangle$ on the de Sitter
(inflating) brane. They found that the Kaluza-Klein contribution is
small relative to the zero-mode contribution even
in the case $H^2\ell^2\gg1$.
Technically, our analysis
is an extension of theirs to a bulk scalar field of arbitrary mass.
The effect of quantum corrections in the brane-world scenario has also
been discussed by Gilkey, Kirsten and Vassilevich \cite{Gilkey:2001mj}.

The paper is organized as follows.
In Sec.~II, following \cite{Himemoto:2001nd,Shiromizu:2000wj},
we formulate the effective gravitational equations
and the Friedmann equation on the brane when a bulk scalar
field is present. In Sec.~III, we discuss in depth the features of
the brane-inflation model of \cite{Himemoto:2001nd} which we use as the background.
We find an indication that the quantum fluctuations of
the bulk scalar field may cause a non-negligible contribution to the
dynamics of the brane when $|m^2|\bar\ell^2\gg 1$.
 In Sec.~IV, we calculate the quantum fluctuations of the bulk scalar field
and evaluate the extra-dimensional effect on the brane.
We find the Kaluza-Klein contribution is indeed
non-negligible if $|m^2|\bar\ell^2\gg1$.
Section V is denoted to summary and discussions.

\section{Formulation}
We write the bulk metric in the $(4+1)$ form as
\begin{equation}
ds^2 =g_{ab}dx^adx^b
= n_an_bdx^adx^b+q_{ab}dx^{a}dx^{b},
\label{eq:metric}
\end{equation}
where $n_a$ is the unit normal to the 4-dimensional
timelike hypersurfaces, one of which is the brane,
and $q_{ab}=g_{ab}-n_an_b$ is the induced metric on
the hypersurfaces. We introduce the coordinates $\{\chi,x^\mu\}$
($\mu=0,1,2,3$) such that $n_adx^a=d\chi$ and assume the brane is
located at $\chi=\chi_0$.

Following the spirit of the Horava-Witten theory
\cite{Horava:1996qa,Horava:1996ma},
we assume the $Z_2$ symmetry with respect to the brane
and that the gravitational equations in the 5-dimensional
spacetime take the following form,
\begin{equation}
G_{ab}+\Lambda_5g_{ab} =
\kappa_5^2[T_{ab}+S_{ab}\delta(\chi-\chi_0)],
\end{equation}
where $G_{ab}$ is the 5-dimensional Einstein tensor and
$T_{ab}$ is the effective energy-momentum tensor
in the 5-dimensional bulk which may consist of higher-order
curvature terms as well as dilatonlike
gravitational scalar fields, and $S_{ab}$ is the energy-momentum tensor
of the matter confined on the brane.

As in the Randall-Sundrum model \cite{Randall:1999ee,Randall:1999vf},
we consider $S_{ab}$ of the form,
\begin{equation}
S_{ab} = -\sigma q_{ab} + \tau_{ab}\,; \quad
\tau_{ab}n^a=0\,,
\label{eq:Sab}
\end{equation}
where $\sigma$ is the brane tension and $\tau_{ab}$ describes
the 4-dimensional matter excitations.
Then the effective gravitational equations on the brane
are found as \cite{Shiromizu:2000wj}
\begin{equation}
{}^{(4)}G_{\mu\nu}+\Lambda_4 q_{\mu\nu}
=\kappa_4^2 (T^{(b)}_{\mu\nu}+\tau_{\mu\nu})
+\kappa_5^4 \pi_{\mu\nu}-E_{\mu\nu} .
\label{eq:Einstein}
\end{equation}
Here
\begin{eqnarray}
\Lambda_4 &=&
\frac{1}{2}(\Lambda_5 + \frac{1}{6}\kappa_5^4 \sigma^2 ) ,
\label{eq:Lambda4def} \\
\kappa_4^2 &=&
\frac{1}{6}\kappa_5^4\sigma ,
\\
\pi_{\mu\nu} &=&
-\frac{1}{4}\tau_{\mu\alpha}\tau_{\nu}^{\alpha}
+\frac{1}{12}\tau\tau_{\mu\nu}
+\frac{1}{8}q_{\mu\nu}\tau_{\alpha\beta}\tau^{\alpha\beta}
-\frac{1}{24}q_{\mu\nu}\tau^2 ,
\label{eq:kappa4def} \\
T^{(b)}_{\mu\nu} &=&
\frac{4}{\sigma\kappa_5^2} \left[
T_{cd}q_{\mu}^c q_{\nu}^d +
\left( T_{cd}n^c n^d - \frac{1}{4}T\right)
q_{\mu\nu} \right],
 \label{eq:Tbdef} \\
E_{\mu\nu} &=&
{}^{(5)}C_{abcd}n^an^cq_{\mu}^bq_{\nu}^d,
\label{eq:Edef}
\end{eqnarray}
where $\Lambda_4$ and $\kappa_4^2$ describe the
4-dimensional cosmological constant and the 4-dimensional
gravitational constant,
respectively, and ${}^{(5)}C_{abcd}$ is the 5-dimensional Weyl tensor.
Note that the projected Weyl tensor
$E_{\mu\nu}$ is traceless by definition.
Note also that the bulk energy-momentum tensor contributes
to a 4-dimensional energy-momentum tensor.
Here and in what follows $q_{\mu\nu}$ denotes the metric on
the brane unless otherwise noted.
We consider a scenario in which the Randall-Sundrum brane is recovered
in the ground state. Thus we assume the brane tension as
\begin{equation}
\sigma=\sigma_c:=\frac{\sqrt{-6\Lambda_5}}{\kappa_5^2}
={6\over\kappa_5^2\ell}\,,
\label{eq:sigmacdef}
\end{equation}
which gives $\Lambda_4=0$.

For the brane cosmology, we assume the spatial homogeneity
and isotropy and express the metric on the brane as
\begin{equation}
ds^2|_{{\rm brane}} =
q_{\mu\nu}dx^{\mu}dx^{\nu} =
-dt^2+a(t)^2\gamma_{ij}dx^idx^j,
\end{equation}
where $\gamma_{ij}$ is the 3-metric of a constant curvature
space with unit or zero curvature, $K=\pm1,0$.
The matter energy-momentum tensor takes
a perfect fluid form on this brane,
\begin{equation}
\tau_{\mu\nu}=
(\rho^{(m)}+P^{(m)})t_{\mu}t_{\nu}+P^{(m)}q_{\mu\nu}\,,
\end{equation}
where $t_{\mu}dx^\mu=-dt$.
With these, the time-time component of Eq.~(\ref{eq:Einstein})
gives the Friedmann equation on the brane,
\begin{eqnarray}
3\left[\left(\frac{\dot{a}}{a}\right)^2+{K\over a^2}\right]
&=&\kappa_4^2\rho^{(m)}+\frac{\kappa_5^4}{12}\rho^{(m)2}
+\kappa_4^2T^{(b)}_{tt}-E_{tt}\,,
\label{eq:Friedmann}
\end{eqnarray}
where the dot denotes the derivative with respect to $t$,
and the trace of the space-space components gives
\begin{eqnarray}
-\left[ 2\frac{\ddot{a}}{a}+\left(\frac{\dot{a}}{a}\right)^2
+{K\over a^2}\right] &=&
\kappa_4^2P^{(m)}+{\kappa_{5}^{4}\over 12} \rho^{(m)}(\rho^{(m)}+2P^{(m)})
+\kappa_4^2{T^{(b)}{}_{i}{}^{i}\over3}-{E_{tt}\over3}\,.
\label{eq:ddota}
\end{eqnarray}
The second equation is not necessary if the
Bianchi identities are taken into account. We give it here for
later convenience.
The Friedmann equation (\ref{eq:Friedmann}) is not closed on the brane
because of the contributions
$T^{(b)}_{tt}$ and $E_{tt}$ which depend on the bulk dynamics.
However, under certain situations as described below,
the scalar field dynamics in the bulk may be solved perturbatively
and $E_{tt}$ may be expressed in terms of the scalar field relatively
easily.

As noted before, the bulk energy-momentum tensor may contain
the contributions from higher-order curvature terms as well
as dilatonlike fields. However, lacking the detailed knowledge
of their explicit forms, we choose the simplest
case of a minimally coupled scalar field,
\begin{equation}
T_{ab}=
\partial_a\phi\partial_b\phi -
g_{ab}\left[ \frac{1}{2}g^{cd}\partial_c\phi\partial_d\phi
             + V(\phi) \right].
\label{eq:Tab}
\end{equation}
As discussed in Ref.~\cite{Himemoto:2001nd}, the assumption
of the $Z_2$-symmetry and
the energy-momentum conservation for the matter field on the brane,
$D_\mu\tau^{\mu\nu}=0$, implies \cite{footnote1}
\begin{equation}
\partial_{\chi}\phi|_{\rm brane} = 0. \label{eq:dphib}
\end{equation}
Hence together with the spatial homogeneity of $\phi$
on the brane, $T^{(b)}_{\mu\nu}$ reduces to
\begin{eqnarray}
T^{(b)}_{\mu\nu}=(\rho^{(b)}+P^{(b)})t_\mu t_\nu+P^{(b)}q_{\mu\nu}\,,
\end{eqnarray}
where
\begin{eqnarray}
\rho^{(b)}&=&T^{(b)}_{tt}=
\frac{3}{\kappa_5^2\sigma_c}\left[\frac{1}{2}\dot{\phi}^2+V(\phi)\right]
={\ell\over2}\left[\frac{1}{2}\dot{\phi}^2+V(\phi)\right],
\label{eq:rhob} \\
P^{(b)}&=&\frac{1}{3}T^{(b)}{}_{i}{}^{i}=
\frac{3}{\kappa_5^2\sigma_c}\left[\frac{5}{6}\dot{\phi}^2-V(\phi)\right]
={\ell\over2}\left[\frac{5}{6}\dot{\phi}^2-V(\phi)\right].
\label{eq:Pb}
\end{eqnarray}

The form of the projected Weyl tensor on the brane
is determined from its traceless nature,
\begin{eqnarray}
E_{\mu\nu}
=\left({4\over3}t_\mu t_\nu+{1\over3}q_{\mu\nu}\right)E_{tt}\,.
\end{eqnarray}
Then, from the 4-dimensional Bianchi identities and the spatial
homogeneity of the brane, we obtain \cite{Shiromizu:2000wj}
\begin{equation}
D^{\nu}E_{\mu\nu}=\kappa_4^2 D^{\nu}T^{(b)}_{\mu\nu}\,,
\label{eq:divE}
\end{equation}
which gives
\begin{equation}
\frac{1}{a^4}\partial_t(a^4E_{tt})
=-\frac{\kappa_5^2}{6}
[\ddot\phi - 4\partial_{\chi}^2\phi + V'(\phi)]
\dot\phi\,.
 \label{eq:divEtt}
\end{equation}
Further, using the scalar field equation at the location
of the brane,
\begin{equation}
\partial_{\chi}^2\phi -
\frac{1}{\sqrt{-q}}\partial_{\mu}(\sqrt{-q}q^{\mu\nu}\partial_{\nu}\phi)
= 0\,,
\label{eq:feqonb}
\end{equation}
and Eq.~(\ref{eq:divEtt}) may be rewritten as
\begin{equation}
\frac{1}{a^4}\partial_t(a^4E_{tt})=
\frac{\kappa_5^2}{2}
\left(\partial_{\chi}^2\phi + \frac{\dot a}{a}\dot\phi \right)
\dot\phi\,.
\end{equation}
Formally integrating the above, we obtain
\begin{equation}
E_{tt} =
\frac{\kappa_5^2}{2a^4}\int^t a^4\dot{\phi}
\left(\partial_{\chi}^2\phi + \frac{\dot{a}}{a}\dot{\phi}\right)dt\,.
\label{eq:Ett}
\end{equation}
Thus $E_{tt}$ is expressed in terms of $\phi$
apart from the integration constant which gives
a term $\propto a^{-4}$ and which should be determined by
the initial condition of the 5-dimensional universe.

\section{Model}
Himemoto and Sasaki proposed a model for brane-world inflation
induced by a bulk scalar field \cite{Himemoto:2001nd}.
Here we adopt their model.
Namely, we assume the potential of the form,
\begin{equation}
V(\phi)=V_0+\frac{1}{2}m^2\phi^2,
\end{equation}
and consider the region $|m^2|\phi^2/2\ll V_0$.
In \cite{Himemoto:2001nd}, it was further assumed that $m^2<0$
in order for a regular solution in
the separable form to exist.
Here we basically follow \cite{Himemoto:2001nd} and assume $m^2<0$
for the moment.
However, one can argue that the singularity may not harm the
brane if the action is still finite \cite{Garriga:2000bq}.
There is yet another argument that supports the admissibility
of the case $m^2>0$ \cite{TaHiSa}.
Hence, we relax this assumption in the end and extend
our analysis to positive values of $m^2$.

Assuming $T_{ab}$ is dominated by $V_0$ at the lowest order
of approximation,
\begin{equation}
T_{ab}\simeq -V_0 g_{ab}\,,
\end{equation}
the effective 5-dimensional cosmological constant
becomes
\begin{equation}
\Lambda_{5,\rm{eff}} =
\Lambda_{5} + \kappa_5^2V_0\,.
\label{eq:Lambda5eff}
\end{equation}
Here we assume that $|\Lambda_5|>\kappa_5^2V_0$ so that the background
space-time is still effectively AdS$_5$.
Then the bulk metric may be written as \cite{Garriga:2000bq}
\begin{eqnarray}
ds^2&=&dr^2+(H\bar{\ell})^2\sinh^2(r/\bar{\ell})\,
ds_{\rm dS}^2\,,
\label{eq:bulkmetric}
\end{eqnarray}
where $ds_{\rm dS}^2$ is the metric of the
4-dimensional de Sitter space with radius
$H^{-1}$ given by
\begin{equation}
H^{-1}=\bar{\ell}\sinh (r_0 /\bar{\ell})\,;
 \quad
\bar{\ell} = \left| \frac{6}{\Lambda_{5,{\rm eff}}} \right|^{1/2}.
\label{eq:Hbarell}
\end{equation}
Note that we have replaced the coordinate $\chi$ in the previous
section by $r$ in accordance with the standard convention, and
the value of $r_0$ (equivalently of $H$)
is arbitrary for the moment.

As noted in the previous section, we assume the brane tension
to be that of the Randall-Sundrum value, given by
Eq.~(\ref{eq:sigmacdef}). Then choosing $r=r_0$ to be
the location of the brane, we have the de Sitter brane
with the metric,
\begin{eqnarray}
ds_{\rm dS}^2=-dt^2+\frac{1}{H^2}e^{2Ht}d\bm{x}^2\,,
\label{eq:dSmetric}
\end{eqnarray}
where we have adopted the spatially flat slicing for simplicity.
Neglecting the matter field excitations on the brane,
the Friedmann equation at the lowest order determines $H$ as
\begin{equation}
H^2=\frac{\kappa_5^2V_0}{6}\,,
 \label{eq:H2eq}
\end{equation}
which in turn determines the location of the brane $r_0$.

The next order correction is determined by solving the
scalar field equation in the bulk,
\begin{equation}
\frac{e^{-3Ht}}{(\bar{\ell}H)^2\sinh^2(r/\bar{\ell})}
\partial_t[e^{3Ht}\partial_t\phi] -
\frac{1}{\sinh^4(r/\bar{\ell})}
\partial_r[\sinh^4(r/\bar{\ell})\partial_r\phi] -
\frac{e^{-2Ht}}{\bar{\ell}^2\sinh^2(r/\bar{\ell})}
\,{}^{(3)}\Delta\phi +
m^2\phi = 0, \label{eq:fieldeq}
\end{equation}
where ${}^{(3)}\Delta$ is the flat Laplacian
with respect to $\bm{x}$.
Assuming that our background solution is in the separable form,
\begin{equation}
\phi(t,r)=\psi(t)u(r).
\end{equation}
the equations for $\psi(t)$ and $u(r)$ become
\begin{eqnarray}
\left[{d^2\over dt^2}+3H{d\over dt}+
\lambda^2 H^2\right]\psi
&=& 0 ,
\label{eq:psieq} \\
\left[{d^2\over dr^2}+
\frac{4}{\bar{\ell}}\coth\left(\frac{r}{\bar{\ell}}\right)
{d\over dr}-
\left(m^2 - \frac{\lambda^2}{\bar{\ell}^2\sinh^2(r/\bar{\ell})}
\right)\right]u &=& 0,
\label{eq:ueq}
\end{eqnarray}
where $\lambda$ is a separation constant \cite{footnote2}.

Equation (\ref{eq:ueq}) is the eigenvalue equation for $\lambda^2$.
The $Z_2$ symmetry implies the boundary condition at the brane,
\begin{eqnarray}
\left.{d\over dr}u\right|_{r=r_0}&=&0\,.
\label{eq:Z2}
\end{eqnarray}
As for the boundary condition at the origin $r=0$,
we impose the regularity of $u$ as in \cite{Himemoto:2001nd},
\begin{eqnarray}
u|_{r=0}=0\,.
\label{eq:regularity}
\end{eqnarray}
Under these conditions, the solution is found as
\begin{eqnarray}
u=u_{\lambda_0}(r) &=&
\frac{{\rm P}_{\nu-1/2}^{-\mu_0}(\cosh(r/\bar{\ell}))}
     {\sinh^{3/2}(r/\bar{\ell})},
\label{eq:u0}
\end{eqnarray}
where $\mu_0:=\sqrt{9/4-\lambda_0^2}$
and $\nu:=\sqrt{m^2\bar{\ell}^2+4}$
with the eigenvalue $\lambda=\lambda_0$ determined by
\begin{equation}
(\nu+2)z_0
{\rm P}_{\nu-1/2}^{-\mu_0}(z_0) =
(\nu+\mu_0+1/2){\rm P}_{\nu+1/2}^{-\mu_0}(z_0) \quad ; \quad
z_0:=\cosh(r_0/\bar{\ell})\,.
\label{eq:eigeneq}
\end{equation}

Provided $|m^2|/H^2\ll 1$, $\lambda_0^2$ is approximately
given by \cite{Himemoto:2001nd}
\begin{eqnarray}
\lambda_0^2=\left\{
\begin{array}{ll}
\displaystyle{m^2\over2H^2}&\quad\mbox{for}\quad|m^2|\bar\ell^2\ll1\,,
\\
~\\
\displaystyle{3m^2\over5H^2}&\quad\mbox{for}\quad|m^2|\bar\ell^2\gg1\,.
\end{array}\right.
\label{eq:lambda0}
\end{eqnarray}
Then the corresponding solution for $\psi$ is given by
\begin{eqnarray}
\psi=\psi_{\lambda_0}(t)
=C\exp\left[\left(\mu_0-{3\over2}\right)Ht\right]
+D\exp\left[\left(-\mu_0-{3\over2}\right)Ht\right]
\label{eq:psi0}
\end{eqnarray}
which behaves as $\propto e^{-\lambda_0^2Ht/3}$
at late times.
Thus slow-roll inflation is realized on the brane
for $|m^2|/H^2\ll 1$, irrespective of the magnitude of
$|m^2|\bar\ell^2$.

Note that the regularity condition (\ref{eq:regularity})
implies that $\lambda^2$ must be nonpositive. This was
the reason why only the case $m^2<0$ was considered
in \cite{Himemoto:2001nd}. However, as mentioned earlier, the singularity
at $r=0$ may be harmless as long as $u_{\lambda_0}$
is square integrable \cite{Garriga:2000bq,TaHiSa}. As we shall see in Sec.~IV,
$u_{\lambda_0}$ is indeed square integrable even for $m^2>0$
and it turns out to be the unique bound-state solution, i.e.,
the zero-mode solution.
Therefore we relax the assumption
$m^2<0$ and consider both signs of $m^2$ in the following.
The eigenvalue $\lambda_0^2$ gives the
effective 4-dimensional mass of the zero mode,
\begin{eqnarray}
m_{\rm eff}^2=\lambda_0^2H^2
=\left\{
\begin{array}{ll}
\displaystyle{m^2\over2}&\quad\mbox{for}\quad|m^2|\bar\ell^2\ll1\,,
\\
~\\
\displaystyle{3m^2\over5}&\quad\mbox{for}\quad|m^2|\bar\ell^2\gg1\,.
\end{array}\right.
\label{eq:zeromass}
\end{eqnarray}

With the above perturbative solution for the scalar field, let us now
consider the correction to the Friedmann equation
on the brane. To see the effect on the dynamics of the brane,
we evaluate the effective energy density and pressure on
the brane. In the present case, Eqs.~(\ref{eq:Friedmann}) and
 (\ref{eq:ddota}) become
\begin{eqnarray}
3\left(\frac{\dot{a}}{a}\right)^2 &=&
\kappa_4^2\left(\bar{\rho}+V_0\right),
\nonumber\\
-\left[ 2\frac{\ddot{a}}{a}+\left(\frac{\dot{a}}{a}\right)^2
 \right] &=&
\kappa_4^2\left(\bar{P}-V_0\right)\,,
\end{eqnarray}
where
\begin{eqnarray}
\bar{\rho} &=&
\frac{\ell}{2}\left[\frac{1}{2}\dot{\phi}^2+{1\over2}m^2\phi^2\right]
-\frac{1}{\kappa_4^2}E_{tt}\,,
\label{eq:rhobar}\\
\bar{P} &=&
\frac{\ell}{2}\left[\frac{5}{6}\dot{\phi}^2-{1\over2}m^2\phi^2\right]
-\frac{1}{3\kappa_4^2}E_{tt}\,.
\label{eq:Pbar}
\end{eqnarray}

Using our solution for the scalar field given by
Eqs.~(\ref{eq:u0}) and (\ref{eq:psi0}),
the integral expression (\ref{eq:Ett}) for $E_{tt}$
can evaluated as
\begin{eqnarray}
E_{tt} &=&
\frac{\kappa_5^2}{2}\frac{\mu_0-3/2}{2\mu_0+1}
\left[m^2-\lambda_0^2H^2+\left(\mu_0-\frac{3}{2}\right)H^2\right]\phi^2 \\
&\simeq&
-\frac{3\kappa_5^2}{8\lambda_0^2H^2}
\left(m^2-\frac{4}{3}\lambda_0^2H^2\right)\dot{\phi}^2,
\label{eq:Ettvalue}
\end{eqnarray}
where we have taken the asymptotic limit $Ht\gg1$
and used the approximation $\mu_0\simeq 3/2-\lambda_0^2/3$.
Inserting the above to Eqs.~(\ref{eq:rhobar})
and (\ref{eq:Pbar}), we find
\begin{eqnarray}
\bar{\rho} &\simeq&
\frac{\ell}{2}\left[
\left(\frac{3m^2}{4\lambda_0^2H^2}-\frac{1}{2}\right)\dot{\phi}^2
+\frac{1}{2}m^2\phi^2 \right],
 \label{eq:rhoeff} \\
\bar{P} &\simeq&
\frac{\ell}{2}\left[
\left(\frac{m^2}{4\lambda_0^2H^2}+\frac{1}{2}\right)\dot{\phi}^2
-\frac{1}{2}m^2\phi^2 \right].
\label{eq:Peff}
\end{eqnarray}

When $|m^2|\bar{\ell}^2\ll |m^2|/H^2\ll 1$, we have
$H^2\bar\ell^2\ll1$. In this case, using the eigenvalue
$\lambda_0^2$ given by Eq.~(\ref{eq:lambda0}), we obtain
\begin{eqnarray}
\bar{\rho} \simeq
\frac{\ell}{2}\left(\dot{\phi}^2+\frac{1}{2}m^2\phi^2\right) =
\frac{1}{2}\dot{\Phi}^2+\frac{1}{2}m_{{\rm eff}}^2\Phi^2,
\nonumber \\
\bar{P} \simeq
\frac{\ell}{2}\left(\dot{\phi}^2-\frac{1}{2}m^2\phi^2\right) =
\frac{1}{2}\dot{\Phi}^2-\frac{1}{2}m_{{\rm eff}}^2\Phi^2,
\label{eq:cllimit}
\end{eqnarray}
where we have introduced the rescaled 4-dimensional
scalar field $\Phi$ defined by
\begin{equation}
\Phi := \sqrt{\ell}\,\phi\,.
\label{eq:4dphi}
\end{equation}
Since $H^2\bar{\ell}^2\ll 1$ in this case, the extra dimension is
sufficiently compact in comparison with the
Hubble radius of the brane.
Therefore the dynamics on the brane on
super Hubble horizon scales is expected to be dominated
by the zero-mode solution and to be essentially the same as
the conventional 4-dimensional theory. The above result
indeed supports this expectation.

In contrast, when $|m^2|\bar{\ell}^2\gg 1$, if the
effective 4-dimensional mass of the scalar field should
be still given by that of the zero-mode solution,
the effective energy density and pressure on the brane
are given by
\begin{eqnarray}
\bar{\rho} &\simeq&
\frac{\ell}{2}\left(\frac{3}{4}\dot{\phi}^2+\frac{1}{2}m^2\phi^2\right)\,,
\nonumber\\
\bar{P} &\simeq&
\frac{\ell}{2}\left(\frac{11}{12}\dot{\phi}^2-\frac{1}{2}m^2\phi^2\right)\,.
\end{eqnarray}
Apparently these do not agree with those given by the zero-mode solution.
If we introduce the rescaled 4-dimensional field for this case as
\begin{equation}
\Phi := \sqrt{\frac{5\ell}{6}}\,\phi\,,
\end{equation}
which minimizes the discrepancy, we have
\begin{eqnarray}
\bar\rho & =&
\frac{1}{2}\dot{\Phi}^2+\frac{1}{2}m_{{\rm eff}}^2\Phi^2
-\frac{1}{20}\dot{\Phi}^2,
\nonumber\\
\bar P &=&
\frac{1}{2}\dot{\Phi}^2-\frac{1}{2}m_{{\rm eff}}^2\Phi^2
+\frac{1}{20}\dot{\Phi}^2.
\end{eqnarray}
It is worth noting that the discrepancy is small even in the limit
$|m^2|\bar\ell^2\to\infty$, provided $\Phi$ is slowly rolling
on the brane.
Nevertheless, this implies that the dynamics on the brane cannot be
described by the zero-mode solution alone even at the
classical level, which in turn suggests that the quantum fluctuations
on the brane may be affected significantly.

Before closing this section, we comment on the possible generality
of our brane-inflation scenario. Since the restriction on the
sign of $m^2$ may be relaxed as discussed above,
it may be said that inflation on the brane
is a natural consequence of the Randall-Sundrum type
brane-world scenario once bulk (gravitational) scalar fields
are taken into account.
The actual issue is not if inflation can occur but how inflation
can end, that is, it is the cosmological constant problem
which should be solved in the brane-world scenario.

\section{Quantum fluctuations}
In this section, we calculate quantum fluctuations of the bulk scalar field.
Kobayashi, Koyama and Soda have recently discussed the effect of
quantum fluctuations of a massless bulk scalar field on the de Sitter brane
\cite{Kobayashi:2001yh}. Here we consider the case of arbitrary mass.

On the bulk background given by the metric (\ref{eq:bulkmetric}),
the fluctuation of the bulk scalar field, $\delta\phi$,
can be expanded as
\begin{equation}
\delta\phi (r,t,\bm{x}) =
\int d\lambda\, u_{\lambda}(r)\phi_{\lambda}(t,\bm{x})\,,
\label{eq:phiexpand}
\end{equation}
where $\phi_{\lambda}$ satisfies the 4-dimensional field equation,
\begin{equation}
\left[-{}^{(4)}\Box + \lambda^2H^2\right]\phi_{\lambda}= 0,
\label{eq:philambda}
\end{equation}
and $u_\lambda$ satisfies the eigenvalue equation,
\begin{eqnarray}
\left[{d^2\over dr^2}
+\frac{4}{\bar{\ell}}\coth\left(\frac{r}{\bar{\ell}}\right)
{d\over dr}
-\left(m^2 - \frac{\lambda^2}{\bar{\ell}^2\sinh^2(r/\bar{\ell})}
\right)\right]u_{\lambda} &=& 0 .
\label{eq:ulambda}
\end{eqnarray}
The d'Alembertian ${}^{(4)}\Box$ is the one with respect to
the de Sitter metric (\ref{eq:dSmetric}). Note that Eq.~(\ref{eq:ulambda})
is the same as Eq.~(\ref{eq:ueq}).

Since $\phi_\lambda$ can be regarded as a 4-dimensional
scalar field with mass $\lambda H$, it can be quantized
following the standard procedure as
\begin{eqnarray}
\phi_{\lambda}(t,\bm{x}) =
\int d^3\bm{k}\,
\left[a_{k\lambda}\psi_{k\lambda}(t)e^{i\bm{k}\cdot\bm{x}}
+ ({\rm H.c.})\right] ,
\end{eqnarray}
where $a_{\bm{k}\lambda}$ and $a_{\bm{k}\lambda}^\dagger$
are the annihilation and creation operators which satisfy
\begin{equation}
[a_{\bm{k}\lambda},\,a^{\dagger}_{\bm{k}'\lambda'}] =
\delta(\bm{k}-\bm{k}')\delta(\lambda - \lambda ') ,
\end{equation}
and $\psi_{k\lambda}$ is the positive frequency function
satisfying
\begin{eqnarray}
\left[{d^2\over dt^2} + 3H{d\over dt}+
(k^2e^{-2Ht}+\lambda^2)H^2\right]\psi_{k\lambda}=0\,;
\quad
\psi_{k\lambda}\dot\psi_{k\lambda}^*
-\dot\psi_{k\lambda}\psi_{k\lambda}^*=iH^3e^{-3Ht}\,.
\label{eq:psilambda}
\end{eqnarray}
Choosing the de Sitter-invariant Euclidean vacuum
as the state to be annihilated by $a_{\bm{k}\lambda}$,
we have
\begin{eqnarray}
\psi_{k\lambda}(t) =
\frac{\sqrt{\pi}}{2}e^{i\pi\mu/2}
H(-\eta)^{3/2}{\rm H}_{\mu}^{(1)}(-k\eta)\,,
\label{eq:psisol}
\end{eqnarray}
where $\eta=-e^{-Ht}$ is the conformal time and
$\mu=\sqrt{9/4-\lambda^2}$.

The eigenvalue equation (\ref{eq:ulambda}) for $u_\lambda(r)$
is solved under the boundary condition,
\begin{equation}
\left.{d\over dr}u_{\lambda}\right|_{r=r_0} = 0\,,
\label{eq:uboundary}
\end{equation}
together with the normalization condition,
\begin{equation}
2(\bar{\ell}H)^{2}
\int_0^{r_0} dr \sinh^{2}(r/\bar{\ell})
u_{\lambda}u_{\lambda '}^{\ast}
= \delta(\lambda - \lambda ').
 \label{eq:unorm}
\end{equation}
This normalization condition ensures the correct
canonical quantization of $\delta\phi$ in the bulk.

Solving Eqs.~(\ref{eq:ulambda}) under these conditions,
we find there are one bound state and an infinite number of continuous modes.
The bound state has the smallest eigenvalue $\lambda_0^2$ and is
the same as the background solution given by Eq.~(\ref{eq:u0})
apart from the normalization. It is given by
\begin{eqnarray}
u_{\lambda_0}(r) &=&
{1\over B H\bar\ell\sinh^{2}(r/\bar{\ell})}
{\rm Q}_{\lambda_0-1/2}^{\nu}( \coth (r/\bar{\ell})),
\label{eq:boundsol} \\
B^2 &=&2
\int_0^{r_0} dr \left(
       \frac{{\rm Q}_{\mu_0-1/2}^{\nu}(\coth(r/\bar{\ell}))}
            {\sinh(r/\bar{\ell})} \right)^2,
\label{eq:boundnorm}
\end{eqnarray}
where $\nu=-\sqrt{m^2\bar\ell^2+4}$,
and $\mu_0 = \sqrt{9/4-\lambda_0^2}$ is determined by
\begin{equation}
\left( \mu-\nu+\frac{1}{2}\right){\rm Q}_{\mu+1/2}^{\nu}
(\coth (r_0/\bar{\ell})) =
\left( \mu - \frac{3}{2}\right) \coth (r_0/\bar{\ell})
{\rm Q}_{\mu-1/2}^{\nu}(\coth ( r_0/\bar{\ell})),
\label{eq:lambda0eq}
\end{equation}
which is equivalent to Eq.~(\ref{eq:eigeneq}) \cite{bateman}.

Rewriting Eq.~(\ref{eq:ulambda}) in the standard Schr\"odinger
form \cite{Himemoto:2001nd}, we find the continuous mode spectrum
for $\lambda>3/2$.
It should be noted that the continuous mass spectrum is
independent of the 5-dimensional mass of the field.
These continuous modes are called the Kaluza-Klein
modes and the existence of them is the main signature of the brane-world.
The Kaluza-Klein mode solutions are given by
\begin{eqnarray}
u_{\lambda}(r) &=&
\frac{1}{N\sqrt{\bar{\ell}}}
\left( \frac{\lambda^2}{\lambda^2 - 9/4} \right)^{\frac{1}{4}}
{1\over H\bar\ell\sinh^{2}(r/\bar{\ell})}
 \nonumber \\
 & & \hspace*{2cm} \times
\left[ {\rm P}_{\mu - 1/2}^{\nu}(\coth(r/\bar{\ell})) -
\alpha_{\lambda}(z_0)
{\rm Q}_{\mu - 1/2}^{\nu}(\coth(r/\bar{\ell})) \right],
\label{eq:thesis6-20}
\end{eqnarray}
where
\begin{eqnarray}
\mu=\sqrt{9/4-\lambda^2}=i\sqrt{\lambda^2-9/4}\,,
\quad
\nu = -\sqrt{m^2\bar{\ell}^2+4}\,,
\end{eqnarray}
and
\begin{eqnarray}
&&N^2=
\left|\frac{\Gamma(\mu)}{\Gamma(\mu-\nu+\frac{1}{2})}\right|^2 +
\left|\frac{\Gamma (-\mu)}{\Gamma (-\mu-\nu+\frac{1}{2})} -
      \pi\alpha_{\lambda}(z_0)
      e^{\nu\pi i}\frac{\Gamma (\mu+\nu+\frac{1}{2})}{\Gamma (\mu+1)}
\right|^2 , \nonumber\\
&&\alpha_{\lambda}(z_0) =
\frac{(\mu-\nu+1/2){\rm P}_{\mu+1/2}^{\nu}(z_0) +
      (3/2-\mu)z_0{\rm P}_{\mu-1/2}^{\nu}(z_0)}
     {(\mu-\nu+1/2){\rm Q}_{\mu+1/2}^{\nu}(z_0) +
      (3/2-\mu)z_0{\rm Q}_{\mu-1/2}^{\nu}(z_0)} \,;
\quad z_0 \equiv \coth(r_0/\bar{\ell}) .
\end{eqnarray}

Assuming the state to be the Euclidean vacuum,
the vacuum expectation value $\langle\delta\phi^2\rangle$
on the brane is given by
\begin{eqnarray}
\langle\delta\phi^2(\eta)\rangle|_{r=r_0}
&=&\int d^3k\, \left[P_0(k;\eta)+\int_{3/2}^\infty d\lambda\,
P(\lambda,k;\eta)\right]\,,
\label{eq:phi2vac}
\end{eqnarray}
where $P_0(k;\eta)$ and $P(\lambda,k;\eta)$
are the power spectra of the zero mode and the Kaluza-Klein
mode,
\begin{eqnarray}
&&P_0(k;\eta)
=|\psi_{k\lambda_0}(\eta)|^2|u_{\lambda_0}(r_0)|^2,
\label{eq:Pzero}\\
&&P(\lambda,k;\eta)
=|\psi_{k\lambda}(\eta)|^2|u_{\lambda}(r_0)|^2.
\label{eq:PKK}
\end{eqnarray}

\begin{figure}[tb]
\begin{center}
\leavevmode
\epsfxsize=16cm
\epsfbox{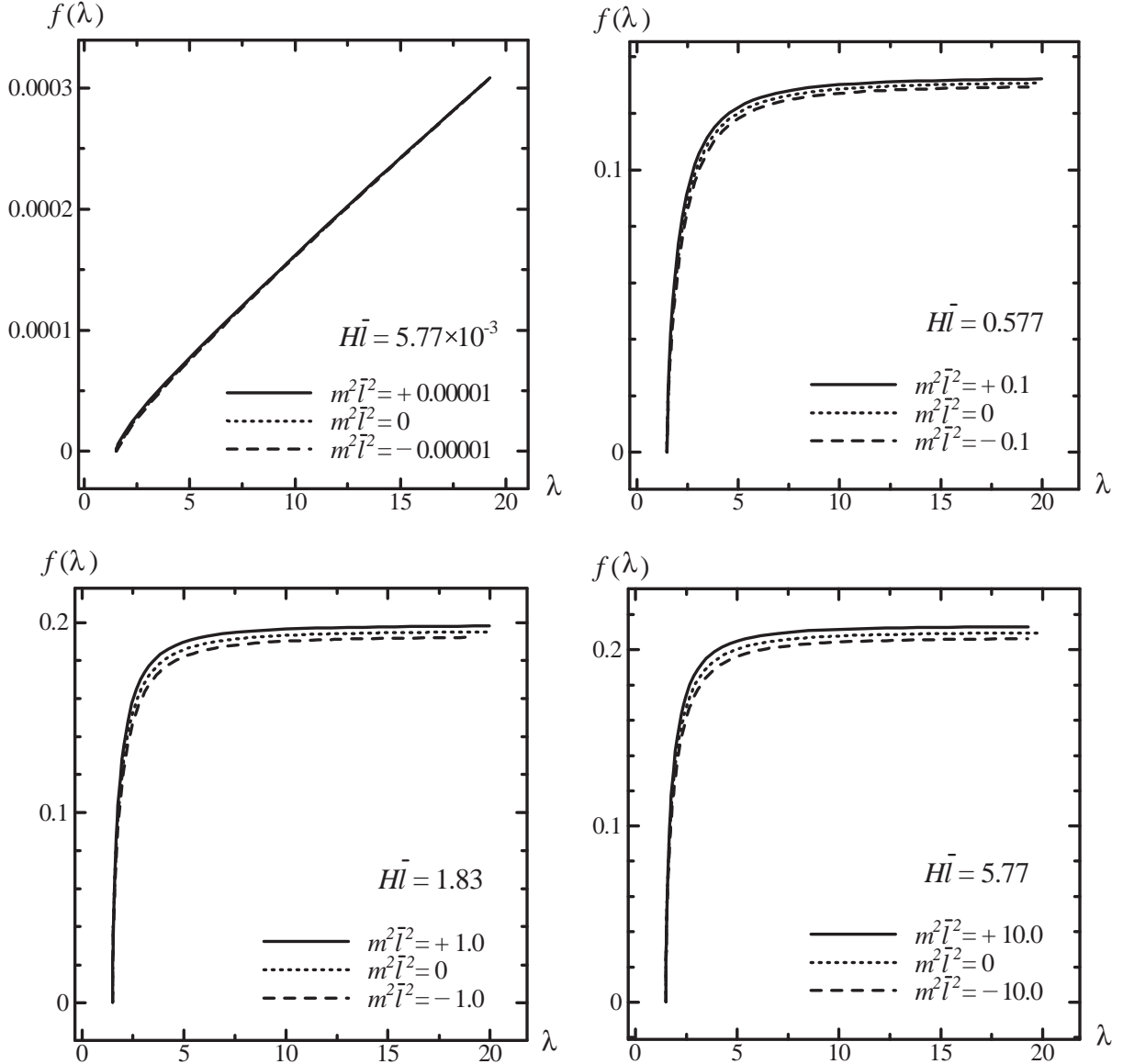}\\
\caption{
 $f(\lambda)\equiv|u_{\lambda}(r_0)|^2/|u_{\lambda_0}(r_0)|^2$
 for various values  of $|m^2|\bar{\ell}^2$.
We set $|m^2|/H^2=0.3$ except for massless cases.
}
\end{center}
\end{figure}

As already noted, the Kaluza-Klein fields $\phi_\lambda$
are independent of the 5-dimensional mass $m$. Thus the difference
between the massless and massive bulk scalar fields appears only
through the difference in the amplitudes of $u_{\lambda}$ on the brane.
To see the contributions of the Kaluza-Klein
modes relative to the zero mode,
we plot $|u_{\lambda}(r_0)|^2/|u_{\lambda_0}(r_0)|^2$ for various
values of $m^2\bar\ell^2$ in Fig.~1.
There we have set $|m^2|/H^2=0.3$ for
all the massive models.  We find the continuous mode spectrum is
rather insensitive to the value of $m^2\bar{\ell}^2$ as long as
$|m^2|/H^2\ll1$.

The bare expectation value (\ref{eq:phi2vac}) is ultraviolet
divergent, and the divergence is cubic now in contrast to
the quadratic divergence in 4-dimensional theory.
As far as the 4-dimensional divergence is concerned,
it may be regularized by cutting off the contributions of the
physical wavenumbers larger than $H$, $-k\eta>1$,
 as in conventional models of inflation.
Then the regularized value may be estimated by
simply evaluating $P_0(k;\eta)$ and $P(\lambda,k;\eta)$
at $-k\eta=1$.
However, the expectation value would still diverge because of
the infinite contribution of the Kaluza-Klein modes.
In the limit $\lambda\to\infty$,
\begin{equation}
P(\lambda,k;\eta)\propto \frac{1}{\lambda}\,,
\label{eq:PKKdiv}
\end{equation}
so the Kaluza-Klein correction diverges logarithmically.

Lacking the relation of $\langle\delta\phi^2\rangle$
to physical observables, we have no principle for how it should
be regularized. Furthermore, the divergencies of the bulk 5-dimensional
theory would require not only local counterterms in the bulk
but also surface counterterms on the brane, and the latter will
generally be nonlocal functionals of the brane metric and
matter fields that may not be absorbed
by the renormalization of 4-dimensional observables.
It is hoped that there exist physically significant observables
that are calculable without fully renormalizing the theory.

Discussions on these issues are, however, out of scope of this paper.
We therefore adopt a simple cutoff regularization
by introducing a cutoff parameter $\lambda_c$
and define the regularized power spectrum,
\begin{equation}
P_{\lambda_c}(k):=
P_0(k)\left[ 1+\int_{3/2}^{\lambda_c}d\lambda\,
{P(\lambda,k)\over P_0(k)}
\right] ,
\end{equation}
where $P_0(k)$ and $P(\lambda,k)$ are $P_0(k;\eta)$
and $P(\lambda,k;\eta)$ evaluated at $-\eta=1/k$, respectively.
Then, in the limit $\lambda_c\to\infty$, we have
\begin{equation}
\Delta_{\lambda_c}:=
\frac{P_{\lambda_c}(k)}{P_0(k)}-1 \to
\alpha+\beta\ln\lambda_c\,.
\label{eq:Deltadef}
\end{equation}
The parameters $\alpha$ and $\beta$ are independent of $\lambda_c$,
and they not only describe the asymptotic behavior of the divergence
but also reflect the cutoff-independent, physical properties of the
regularized spectrum.
Table~I gives the parameters $\alpha$
and $\beta$, and the Kaluza-Klein correction $\Delta_{\lambda_c}$
for several choices of the cutoff parameter $\lambda_c$,
for various cases of the 5-dimensional mass but with the same
value of $|m^2|/H^2=0.3$.
Since the Hubble scale $H$ is the only natural scale we have on the
brane, the value of $\lambda_c$ is taken to be not too greater than $H$.

{}From Table~I, we see that
both $\alpha$ and $\beta$ decrease as $|m^2|\bar\ell^2$ becomes small.
Hence we deduce that the Kaluza-Klein correction vanishes in the limit
$|m^2|\bar{\ell}^2\to 0$. This is consistent with our intuition
that the 5-dimensional theory reduces to the conventional
4-dimensional theory for $\bar{\ell}\ll 1/H$.
The Kaluza-Klein correction becomes large as $|m^2|\bar{\ell}^2$
increases. But instead of growing indefinitely, it seems to be
saturated at a small finite value.
This is the quantum version of the fact we found
for the classical background in Sec.~III, that is, the zero-mode
dominance of the classical solution even in the limit
$|m^2|\bar\ell^2\to\infty$.
Nevertheless, in contrast to the case of the classical background,
the quantum Kaluza-Klein correction is non-negligible,
of the order of about $10\%$.
\begin{table}[htb]
\arraycolsep=6mm
\begin{center}
$$
 \begin{array}{c|cc|ccc} \hline
\vphantom{\displaystyle{A\over B}}
  m^2\bar{\ell}^2 & \alpha & \beta
      & \Delta_{\lambda_c=2} & \Delta_{\lambda_c=5}
      & \Delta_{\lambda_c=10} \\ \hline
  -10.0     & -0.058 & 0.096 & 0.016  & 0.097 & 0.16 \\
  -1.0      & -0.056 & 0.089 & 0.014  & 0.089 & 0.16 \\
  -0.1      & -0.046 & 0.060 & 0.0070 & 0.053 & 0.093 \\
  -10^{-5}  & -0.003 & 7.5\times 10^{-4}
  & 1.5\times10^{-6}   & 2.1\times10^{-5} & 5.9\times10^{-5} \\
  +10^{-5}  & -0.003 & 8.5\times 10^{-4}
  & 2.1\times10^{-6}   & 2.5\times10^{-5} & 6.8\times10^{-5} \\
  +0.1        & -0.051 & 0.071 & 0.0093 & 0.065 & 0.11 \\
  +1.0        & -0.062 & 0.11  & 0.019  & 0.11  & 0.18 \\
  +10.0       & -0.064 & 0.12  & 0.021  & 0.12  & 0.20 \\ \hline
 \end{array}
$$
\caption{
The Kaluza-Klein correction to the power spectrum of
$\langle\delta\phi^2\rangle$ on the brane for various values of
$m^2\bar\ell^2$. The parameter $|m^2|/H^2$ is set to 0.3 for all the
entries. See Eq.~(\protect\ref{eq:Deltadef}) for the definitions
of $\alpha$, $\beta$ and $\Delta_{\lambda_c}$.
}
\end{center}
\end{table}

Before concluding this section,
it may be worth mentioning the following observation.
In the case of $m^2<0$, one might expect instability of some modes to
appear if $|m^2|\bar\ell^2\gg1$, that is, when the bulk curvature
is negligible on the Compton length scale of the field.
But our result that the Kaluza-Klein correction always remains small
implies the absence of such instability.
This may be understood as follows. If we transform the bulk metric
(\ref{eq:bulkmetric}) to a static chart \cite{Kraus:1999it},
\begin{equation}
ds^2 =
-\frac{R^2}{\bar{\ell}^2}dT^2
+ \frac{\bar{\ell}^2}{R^2}dR^2 + R^2d\bm{x}^2.
\end{equation}
the location of the brane can be parametrized as
$(T,R)=(T(\tau),R(\tau))$ where
\begin{equation}
R(\tau)=\frac{1}{H}e^{H\tau} , \quad
T(\tau)=\bar{\ell}\sqrt{1+(H\bar{\ell})^2}\,e^{-H\tau} .
\end{equation}
Since the growth rate of instability would be $O(|m|)$,
while the brane radius increases exponentially at the rate
$H$, the slow-roll condition $|m^2|/H^2 \ll 1$ implies that
the field in the bulk is stretched so fast that the instability has
no time to grow.

\section{Summary}
Based on a brane-inflation model {\it \`a la\/}
Randall-Sundrum brane world \cite{Randall:1999ee,Randall:1999vf}
proposed by Himemoto and Sasaki \cite{Himemoto:2001nd},
we have investigated the quantum fluctuations of a massive
bulk scalar field $\phi$ with mass $m$
which drives slow-roll inflation on the brane.
In this model, the potential of the scalar field $V(\phi)$ ($>0$)
modifies the curvature radius of the 5-dimensional anti-de Sitter
space from the Randall-Sundrum value $\ell$ to $\bar\ell$ ($>\ell$),
and inflation occurs on the brane with the expansion rate $H$
proportional to $\sqrt{V(\phi)}$.

We have calculated the power spectrum of the expectation value
$\langle\delta\phi^2\rangle$ on the brane. Provided $|m^2|/H^2\ll1$
which ensures slow-roll inflation of the classical background,
we have found that the contribution of the Kaluza-Klein modes,
which features the existence of the extra dimension, is much smaller
than that of the zero mode if $|m^2|\bar\ell^2\ll 1$, whereas it is
small but non-negligible if $|m^2|\bar\ell^2\gg 1$.
The Kaluza-Klein contribution in the latter case has been found
to be of the order of $10\%$. Although $\langle\delta\phi^2\rangle$
does not describe any observable quantity in our model,
our result indicates the importance of
the Kaluza-Klein contributions to physical observables
when $|m^2|\bar\ell^2\gg1$. In particular, the subsequent
cosmological evolution may be substantially affected.

In our analysis, we had to introduce a cutoff parameter to
regularize the logarithmic divergence of the power spectrum
which arises from integral over the infinite number of the
Kaluza-Klein modes, because of the lack of
physical significance in $\langle\delta\phi^2\rangle$ itself
as mentioned above.
Therefore, in order to verify our conclusion in a more rigorous
and quantitative manner, it is necessary to evaluate
quantities directly related to observables instead of
$\langle\delta\phi^2\rangle$. One such is
the cosmological density perturbations in our brane-inflation
model. This issue is left for future study.

\acknowledgements

We would like to thank U. Gen, T. Tanaka, and J. Yokoyama
for stimulating discussions.
This work was supported in part by the Yamada Science Foundation.


\begin{thebibliography}{99}
\bibitem{footnote1}
It is possible to introduce a coupling between $\phi$ and the
matter field on the brane. However, we do not consider such a case
in the present paper.
\end{thebibliography}
\end{document}